# Scalability of Voltage-Controlled Filamentary and Nanometallic Resistance Memories

Yang Lu[‡], Jong Ho Lee[‡], and I-Wei Chen*

[‡]These two authors contribute equally to this work.

Department of Materials Science and Engineering, University of Pennsylvania, Philadelphia, PA 19104-6272, USA

Much effort has been devoted to device and materials engineering to realize nanoscale resistance random access memory (RRAM) for practical applications, but there still lacks a rational physical basis to be relied on to design scalable devices spanning many length scales. In particular, the critical switching criterion is not clear for RRAM devices in which resistance changes are limited to localized nanoscale filaments that experience concentrated heat, electric current and field. Here, we demonstrate voltage-controlled resistance switching for macro and nano devices in both filamentary RRAM and nanometallic RRAM, the latter switches uniformly and does not require forming. As a result, using a constant current density as the compliance, we have achieved area-scalability for the low resistance state of the filamentary RRAM, and for both the low and high resistance states of the nanometallic RRAM. This finding will help design area-scalable RRAM at the nanoscale.

**Introduction**

Nanoscale nonvolatile Resistance Random Access Memory (RRAM) stores information as two distinct resistance states—the high resistance state (HRS, "0") and the low resistance state (LRS, "1")—in an insulating thin film between two electrodes.[1,2] The conceptually simple construct has enormous potentials for use in stand-alone or embedded memory as well as in neuromorphic computing.[3-5] The most commonly studied RRAM is filamentary, in which resistance is switched in localized conducting nanofilaments rich in defects/anion-vacancies (as in valence-change memory, VCM[6]) or metal atoms and cations (as in electrochemical metallization memory, ECM[7]). Size-independent HRS and LRS resistances are often realized in filamentary RRAM,[8-11] such as in TiN/Hf/HfO$_x$/TiN (VCM) devices.[10] But area-scalable HRS and LRS has also been reported in other filamentary RRAM, e.g., in Pt/NiO:Ti/Pt[12] and TiN/Al$_2$O$_3$/TiO$_2$/TiN devices,[13] and why they behave differently is not known. In this regard, the so-called "universal set/reset characteristics" for metal-oxide RRAM,[14,15] which refers to the inverse correlation between the LRS resistance ($R_{LRS}$) and the compliance current ($I_{cc}$) used to switch-on the LRS: $R_{LRS}=b/I_{cc}$,[16-18] is relevant. This is because if the correlation is followed, then a compliance of constant-current-density, $j_{cc}= I_{cc}/A$, $A$ being the device area, will lead to $R_{LRS}\sim 1/A$,[12] thus area scaling. But the origin of this correlation is not apparent from the consensus filamentary mechanism that involves Joule heating, ion migration and electron hopping/emission/tunneling, which are all highly non-linear processes sensitive to the complex temperature-current interplay during on-switching.[15,19]

Beyond filamentary RRAM, there is another class of RRAM that apparently switches uniformly.[20,21] The best evidence for its uniform switching of these so-called nanometallic RRAM came from fracture tests,[22] in which a RRAM is physically severed and the resistances of the two severed halves are individually measured. When a nanometallic RRAM such as those made of Mo/SiN$_{4/3}$:Pt/Pt was tested,[23] uniform resistivity was found in both halves for both LRS and HRS. In contrast, when a filamentary RRAM made of Ti/HfO$_2$/Pt was tested,[11] one severed halve inherited the resistance of the parent, intact LRS, while the other halve was orders of magnitude more resistive. Another clear contrast of the two types of RRAM came from the capacitance measurement: The capacitance of the filamentary Ti/HfO$_2$/Pt RRAM is independent of the resistance state (e.g., HRS vs. LRS) consistent with the picture of very thin filaments, but the capacitance of the nanometallic Mo/SiN$_{4/3}$:Pt/Pt decreases with decreasing device resistance consistent with a uniform switching picture.[23,24] It would seem natural to expect the resistance of a uniform switching RRAM like the nanometallic RRAM should obey area scaling in both resistance states.

In this paper, we will examine the interplay of switching voltage and current compliance, and see how it impacts the area-scaling laws of the LRS and HRS resistance in these two prototypical RRAMs down to the nano scale. Hopefully, it will shed new light on the "universal set/reset characteristics," which we believe stems from voltage-controlled switching physics, at all length scales. The issue of area-scaling is especially critical for utilizing nano arrays of RRAM, each of which must be paired with a transistor or a selector or else the LRS will provide sneak paths that invalidate the memory array.[25] If the transistor/selector can only apply a constant current density, then any non-area-scaling RRAM will eventually run out of the driving current when it is downsized. Therefore, our finding will help guide design of RRAM nanodevices.

**Results and Discussion**

Compliance effect vs. voltage-controlled switching

For the filamentary RRAM, we fabricated Ti/HfO$_2$/Pt devices by placing a 10 nm-thick atomic-layer-deposited HfO$_2$ film between a Pt bottom electrode and a Ti top electrode of various sizes. After being formed by a negative voltage, without compliance, to HRS,[11] it was switched-on at a positive voltage and switched-off at a negative voltage as shown in **Figure 1a**. (Here, positive voltage causes a current to flow from Ti electrode to Pt electrode.) For the nanometallic RRAM, we similarly fabricated Mo/SiN$_{4/3}$:9.3%Pt/Pt devices with a sputtered amorphous Si$_3$N$_4$-Pt alloyed film sandwiched between Mo and Pt.[23] It did not require forming and could be bipolarly switched in a similar way (**Figure 1b**).

These devices share some additional common features when the sweeping voltage changes. First, the LRS is progressively lowered by increasing the current (**Figure S1**). In fact, the LRS can become so conducting that the device behaves like a metallic one showing a weak but positive temperature coefficient of resistance.[24,26] This tunability is a clear indication that compliance is important in controlling LRS. Second, as the LRS decreases, the voltage required to switch-off the LRS to HRS increases (**Figure S1**), which implies voltage sharing between the switching element and the load (from electrode, interface, line resistance)—such sharing is increasingly in load's favor when the LRS becomes more conductive. If we envision an equivalent circuit of a switchable resistance $R_S$ in series with a load resistance $R_{load}$ (**Figure 2a**), and let the voltage on $R_S$ be $V^*$, then the applied voltage $V_{app}$ is

$$V_{app} = V^* + I R_{load} \qquad (1)$$

where $I$ is the current of the circuit. As shown in **Figure 2b**, the $V_{app}$-$I$ data in on- and off-switching of both RRAMs in different sizes all fall on a straight-line, giving the same constant $V^* = +0.99$ V and constant $R_{load} = 758$ Ω for the Ti/HfO$_2$/Pt device, and $V^* = +0.97$ V and $R_{load} = 260$ Ω for the Mo/SiN$_{4/3}$:Pt/Pt device. This is strong evidence that on-switching and off-switching in both devices are voltage-controlled. Third, since the load in these two devices is primarily due to the spreading resistance of the bottom electrode, which has a very weak, logarithmic dependence on area, $R_{load}$ is insensitive to area.[11,27] (More generally, $R_{load}$ contains line resistance, which is also insensitive to area in many cases.) Therefore, defining $R_{LRS}$ as $V_{app}/I$ thus $R_{LRS} = V^*/I + R_{load}$ from Eq. (1), we find the $R_{LRS}$ of both RRAMs will naturally approach an area-insensitive asymptote $R_{load}$ when a sufficiently large switch-on current drives $R_S \ll R_{load}$.[28,29] This is the case for the LRS data in **Figure S2**, which vary within a factor of 2 for devices of different areas. In comparison, since $R_S \gg R_{load}$ in the HRS, the HRS is asymptotically $R_S$, which is insensitive to the area in the Ti/HfO$_2$/Pt device (**Figure S2a**) but is area-dependent in the Mo/SiN$_{4/3}$:Pt/Pt device (**Figure S2b**). Therefore, both RRAMs share the same features in their resistance/current/voltage/area interdependence, but their HRS has different area dependence: The HRS of the Ti/HfO$_2$/Pt memory is area independent but the HRS of the Mo/SiN$_{4/3}$:Pt/Pt memory follows area scaling.

The above findings are supported by further statistical data shown in the Weibull plots in **Figure 2c-f.** These data were obtained from DC/AC (100 ns) testing of devices of 2.5 μm to 10 μm in size and included a higher Pt composition (15.5%) for the Mo/SiN$_{4/3}$:Pt/Pt device. For both RRAMs, $V^*$ are similarly distributed around ±1 V regardless of cell area (**Figure 2c-d**). The resistance of the Mo/SiN$_{4/3}$:Pt/Pt device (**Figure 2f**) is tightly distributed and shows area scaling for HRS but area independence for LRS; for the Ti/HfO$_2$/Pt device (**Figure 2e**), the scatter is much larger, but clearly both LRS and HRS are area independent. Overall, $V^*$ is always at around ±1 V, and the above finding on area dependency/independency of HRS and LRS holds for over 5 orders of magnitude in device area as summarized in **Figure 2g-h**.

We have further analyzed the $V_{app}$-$I$ data of several oxide VCM RRAMs in the literature (**Table 1**), covering all the ones that are regarded as the most promising candidates for nanoscale non-volatile memories and neuromorphic synapses.[30-40] These $V_{app}$-$I$ (or $R$-$I$, where $R=V_{app}/I$) data were originally plotted in the log-log or semi-log scale, which cannot reveal the linear relation in Eq. (1). So we have replotted the same data in the linear scale. As shown in **Table 1**, regardless of the current levels (from mA, to sub-μA), size (from $100 \times 100$ μm$^2$ to $50 \times 50$ nm$^2$), device configuration (from vertical MIM, to crossbar, to 3D stack), and switching materials (from TiO$_x$, Cu$_x$O, ZrO$_2$, HfO$_2$, to NiO), all the on/off-switching $V_{app}$-$I$ data follow a straight line giving an intercept $V^*$ and a slope $R_{load}$. Therefore, these filamentary devices apparently switch by a voltage-controlled mechanism. This holds for not only large devices but also nano devices. It follows that the compliance effect seen in these devices can be similarly explained by voltage-switching mechanism.

Constant current-density compliance

The above results provide a straightforward explanation of the "universal set/reset characteristics": Since $V^*$ is a constant and equals to the product of the compliance current $I_{cc}$ ($I_{cc}=I$) and $R_s$, it follows that one can tune $R_s$ by varying $I_{cc}$, hence tune LRS to a large extent. In particular, if $R_s$ is so tuned that $R_s \gg R_{load}$, then $R_{LRS} = V^*/I + R_{load} \cong V^*/I$, which has the same form as the universal characteristics of $R_{LRS}=b/I_{cc}$. In the following, we will explore the possibility of $R_s$-tuning in the $R_S \gg R_{load}$ limit in the two model RRAMs by performing experiments under controlled compliance. The area scaling in such limit will be assessed.

For this purpose, we will interchangeably use two methods to supplement the $R_{load}$: Adding an external serial transistor, or adding an external serial standard resistor. (Single transistor's current vs. voltage, i.e., $I$-$V_{DS}$ curve, is shown in **Figure 3a**). When a transistor was used, as shown in **Figure 3b**, it was connected in series with the RRAM during on-switching (the positive sweep) in the so-called one-transistor-one-resistor (1T-1R) configuration. But to focus on the current/resistance of on-switching and to avoid any effect of transistor's resistance on the reading of off-switching, we disconnected the transistor from the circuit before continuing with the negative sweep.

The obtained $I$-$V$ and $R$-$V$ curves are depicted in **Figure 3c** and **3d** for both devices, with on-switching and off-switching indicated, and additional data are shown in **Figure S3** for different device areas. All of them verify that the on-switching current is indeed limited by the saturation current of the transistor once the source-to-drain voltage of the transistor exceeds the threshold voltage; this is unlike the case of **Figure 1a-b** and **Figure S1-2** in which the on-switching current continuously increases with the sweeping voltage. By tuning the gate voltage of the transistor to adjust the saturation current according to **Figure 3a**, we can then tune the value of LRS. A serially connected resistor was found to yield essentially the same results as shown in **Figure S4,** albeit the voltage required for switching increases with the load because, unlike the transistor, the standard resistor is sharing the voltage.

A summary of resistance ($R$), capacitance ($C$), current ($I$) and voltage ($V^*$) data including their area dependence is given in **Table 2**, along with several other device characteristics of the two RRAMs mentioned above. (Data supporting each entry in the table are referenced in the entry; they are plotted in **Figure 2, Figure 4**, **Figure S5**, and Refs. 11, 23, 24 and 26.) For both RRAMs, the LRS resistance is area insensitive under a constant-current compliance $I_{cc}$ (**Figure**

**S5c-d**), but it follows the $1/A$ scaling under a constant-current-density compliance $j_{cc}=I_{cc}/A$, the actual data showing $R_{LRS}\sim A^{-0.9\pm0.1}$ for Ti/HfO$_2$/Pt RRAM and $R_{LRS}\sim A^{-0.85\pm0.02}$ for Mo/SiN$_{4/3}$:Pt/Pt RRAM (**Figure S5a-b**). Meanwhile, the HRS always stays constant for Ti/HfO$_2$/Pt RRAM but follows $1/A$ scaling for Mo/SiN$_{4/3}$:Pt/Pt RRAM regardless of the type of compliance (**Figure S5**). Note that in obtaining HRS, a maximum negative sweep voltage of −4 V was used in these tests, which loaded the switched-off device to a voltage considerably more negative than the off-switching voltage, thus ensuring complete switching.

A more detailed study of how voltage, current, current density, etc., affect resistance switching provides further data that directly supports a constant $V^*$. As plotted in **Figure 4a-b**, although the transistor was not included during off-switching, the off-switching current is almost identical to the on-switching current set by the transistor compliance in both RRAMs, which implies the magnitude of $V^*$ is almost the same despite the opposite sign in bipolar switching. In the Mo/SiN$_{4/3}$:Pt/Pt devices, this relation holds when the switching current varies over 4 orders of magnitude from 0.1 μA to 1 mA. (Ti/HfO$_2$/Pt RRAMs were formed to HRS by a negative bias, so the subsequent negative off-switching current has nothing to do with the forming current.) Since the value of $R_S$ in these compliance tests is mostly much larger than that of $R_{load}$, the voltages on $R_S$ can be directly read from the $R$-$V$ switching curves in **Figure S3** in nearly all cases—after subtracting the transistor voltage. For both on-switching and off-switching in either RRAM and over 6 orders of device area, it is near ±1 V in **Figure 4c-d** regardless of on/off current or current density, including tests that applied no current compliance. (The latter data were previously shown in **Figure 2c-d**.) The constant on-switching voltage in Mo/SiN$_{4/3}$:Pt/Pt RRAM was also directly verified using the current-force measurement shown in **Figure S6**: As the current is swept from 0 to 70 μA, the voltage stays constant around 1 V while on-switching proceeds to an increasing extent, causing resistance to decrease continuously. Given the fact that on-switching voltage is invariant and about the same as the off-switching voltage in all cases, the data of **Figure 4a-b** imply $R_S$ remains constant in the duration between on-switching and off-switching, thus excluding possible overshoot effect in the current-limited on-switching.

The various scaling behavior can be easily understood using the simplified $I$-$V_s$ curve in **Figure 4e** that recaps the "universal set/reset characteristics" for an ideal voltage-controlled switching device in the equivalent circuit of **Figure 2a**. At a characteristic voltage $V^*$ across the switching element $R_S$, the current abruptly switches on and takes off. Conversely, at a characteristic voltage $-V^*$, the current abruptly switches off and vanishes. (See **Figure S7** for several examples in both $I$-$V_s$ and $I$-$V_{app}$.) Viewing hysteretic switching as a first-order phase transition, we may regard the current in **Figure 4e** as a measure of the *extent* of the transition, and the voltage a measure of the *driving force*. It is then clear that the phase transition ensues at a critical driving force, but the extent of the transition is dependent on the constraint, such as the current compliance imposed on the memory circuit. The characteristic built-in load, $R_{load}$, defines a characteristic current $I_o=V^*/R_{load}$, which specifies, within a factor two or so, the maximum current that can pass through the bare device when the device is biased by $V^*$ and $R_S=0$, which is asymptotically achieved under a large current. This case applies with the current typically reaching several mA or more and $R_S \ll R_{load}$ when there is no intentionally supplied additional compliance to limit the current. We may term this the *extrinsic* regime, in which LRS is essentially $R_{load}$ and thus area-independent. In contrast, in the *intrinsic* regime, an additional compliance $I_{cc} \ll I_o$ is provided to limit the current, which renders the built-in load unimportant ($R_S \gg R_{load}$). The LRS is now essentially $R_{LRS} \sim R_S = V^*/I_{cc}$. As $V^*$ is found to be independent of the device area, it follows that if the compliance is set at a constant current, then the LRS is area-independent. But if the compliance is set at a constant current density $j_{cc}$, then one obtains $R_{LRS}$ ~$1/A$. This explains the LRS scaling in **Table 2** and **Figure S5**.

Although the $R_s$-tunability of LRS holds for both RRAMs, the Ti/HfO$_2$/Pt device suffers from a decreasing on-off resistance ratio at small sizes because its HRS is area independent. In contrast, since the Mo/SiN$_{4/3}$:Pt/Pt RRAM has an area scalable HRS, it can maintain a constant on-off resistance ratio at least down to sub-micron sizes as shown in **Figure 4f**, where an LRS > 1 MΩ, a switching current < 1 µA, and an off/on ratio of 100 are achieved. This is illustrated by the line in **Figure 4f**, which predicts $V^*/I_{cc}$ with $V^*$ of 1 V, which is in reasonable agreement with the magnitude of the LRS. (A similar fitting for large devices was shown in **Figure S5**.) The 2.5 µm Mo/SiN$_{4/3}$:Pt/Pt device has highly reproducible (100) cycle-to-cycle $R$-$V$ switching curves shown in **Figure S8**, which is consistent with the Weibull statistics in **Figure 2c-f**. (The switching data of this device were included in **Figure 4b** and **Figure 4d**.)

**Further Discussions**

A few more points on area-scaling and voltage-controlled switching follow.

(i) In a nanometallic RRAM that switches uniformly, a constant-current-density compliance $j_{cc}$ apparently causes a constant resistivity throughout the device area $A$, which gives LRS~$1/A$. However, in a filamentary RRAM that is likely to have the same filament size regardless of device area $A$, a constant-current-density compliance $j_{cc}$ must cause a lower resistivity in the filament to give LRS~$1/A$. Therefore, the observation of LRS~$1/A$ tells nothing about whether switching is uniform or filamentary. One must use other tests, such as mechanical fracture tests,[24] to more directly probe the spatial distribution of conducting paths.

(ii) In principle, full scalability is possible in filamentary RRAM if a constant areal density of filaments can be maintained, which might be achieved under special forming conditions, as putatively in the case in the $TiO_2$ devices.[13] However, since forming involves field-induced dielectric breakdown and is inherently a process of stochastic instability,[41,42] achieving a constant filament density in a nanosize device cannot always be assured. (In our experience with $Ti/HfO_2/Pt$ devices, negative forming to HRS always resulted in a single-filament regardless of device area, even though additional incipient filaments apparently also formed according to the fracture experiment.[22]) In the extreme nanoscale limit when the device size becomes comparable to the cross-section of a nanofilament, then perhaps the filament size is dictated by the device size, which could allow effectively uniform switching and full scalability.[43] Another possibility is to invoke an alternative mechanism of oxygen-vacancy conduction, as suggested by Govoreanu et al.[13]

(iii) The simple assumption of voltage-controlled switching is consistent with all the data in this work as well as those of the most advanced RRAM nanodevices in the literatures. It can also explain the results under at least three loading conditions: without an external load, with a current compliance, and with a current density compliance. Naturally, voltage-controlled transitions suggest energy-controlled switching, which is most easily understood if it is an electron process.[23] In $Mo/SiN_{3/4}$:Pt/Pt RRAM, the $R_S$ decrease is mediated by voltage-controlled charge detrapping, which lowers Coulomb repulsion for charged carriers. However, in $Ti/HfO_2/Pt$ RRAM, the $R_S$ decrease has been commonly associated with an increase in the cross section of the filament,[44] which is probably not a voltage-controlled electron process. So the origin of voltage control in filamentary VCM remains in question. Regardless of the mechanism, however, we believe our finding of voltage control at the nanoscale for two broad types of RRAMs will prove conceptually and practically valuable for engineering rationally designed RRAM nanodevices.

**Conclusions**

We may now conclude:

1. Under various current levels and in different device sizes down to the nanoscale, bipolar resistance switching occurs in both filamentary and nanometallic RRAM when a characteristic voltage is supplied to the essential switching element in the film, i.e., their switching is voltage-controlled, which explains the universal correlations between current and low-resistance value in all types of RRAM devices, including nanodevices.

2. When a device-area-proportional current is applied to the switching element in a single-filament RRAM, it can achieve area scaling for LRS but not for HRS. Therefore, the RRAM suffers from a decreasing resistance ratio as the device size shrinks. To overcome this problem, a constant filament density must be maintained in the devices, presumably by advanced control of the forming process.

3. When a device-area-proportional current is applied to the switching element in a nanometallic RRAM, it acquires area-scaling LRS and HRS with the same HRS-to-LRS ratio. An on-current less than 1 μA, an LRS greater than 1 MΩ, and a resistance ratio exceeding 100 have been achieved.

**Methods**

*Device Fabrication*: Filamentary RRAM (Ti/HfO$_x$/Pt) and nanometallic RRAM (Mo/SiN$_{4/3}$:Pt/Pt) devices were fabricated on a substrate of thermal-oxide-coated 100 *p*-type silicon single crystal. Two sets of devices were fabricated. (a) For large sized ones, a 20 nm Mo (NM-RRAM) and 15 nm Ti (for NF-RRAM) were first deposited as bottom electrode, using DC sputtering and e-beam evaporation, respectively. For the switching layer in nanometallic RRAM, a 7.5 nm Si$_3$N$_4$:Pt film was deposited by co-sputtering Si$_3$N$_4$ and Pt using RF power. Likewise, in filamentary RRAM, a 10 nm HfO$_x$ layer was deposited by atomic layer deposition (ALD) at 250$^o$C using tetrakis (dimethylamido) hafnium (HFDMA) precursor and H$_2$O. Finally, a 40 nm Pt top electrode was deposited by RF sputtering through a shadow mask, which defined devices with a radius from 20 μm to 200 μm. (b) For smaller sized nanometallic RRAM devices, a 2 nm Cr bonding layer followed by a 15 nm Pt bottom-electrode layer was deposited on a thermally oxidized silicon wafer using e-beam evaporation and DC sputtering, respectively. Next, a 100 nm-thick SiO$_2$ coating was deposited by plasma enhanced chemical vapor deposition (PECVD) at 300$^o$C using SiH$_4$ and N$_2$O as a silicon and oxygen source, respectively. Sub-μm through holes were patterned onto SiO$_2$ using e-beam lithography, followed by wet etching using buffered oxide etch (BOE). Conventional ultraviolet lithography was then used to pattern 80 μm square holes with pre-patterned sub-μm holes at the center of the 80 μm holes, and a 2.5 nm Si$_3$N$_4$:Pt film followed by a 40 nm Mo top electrode were sputter deposited. Lastly, a final lift-off was performed to pattern the Si$_3$N$_4$:Pt and Mo top films.

*Electrical measurement*: Electrical properties were measured using several electrical meters on a Signatone S-1160 probe station. In a typical test configuration, a bias voltage was applied to the bottom Mo electrode while the bottom contact was grounded. Here, the positive voltage is defined as the voltage that causes a current to flow from Mo or Ti electrode to Pt. Forming of filamentary RRAM into a high resistance state was performed using a negative voltage without any compliance. During on-switching of the 1T1R configuration, a transistor (LND150, n-channel depletion-mode MOSFET, Supertex Inc.) or standard resistor was serially connected with the RRAM device. It was then disconnected from the circuit in off-switching. DC current–voltage (*I-V*) characteristics were examined using a semiconductor parameter analyzer (SPA, Keithley 237). Using the same analyzer, other voltage sweeps and current sweeps were performed in the standard way, the latter typically from 0 to 70 µA under a positive bias.

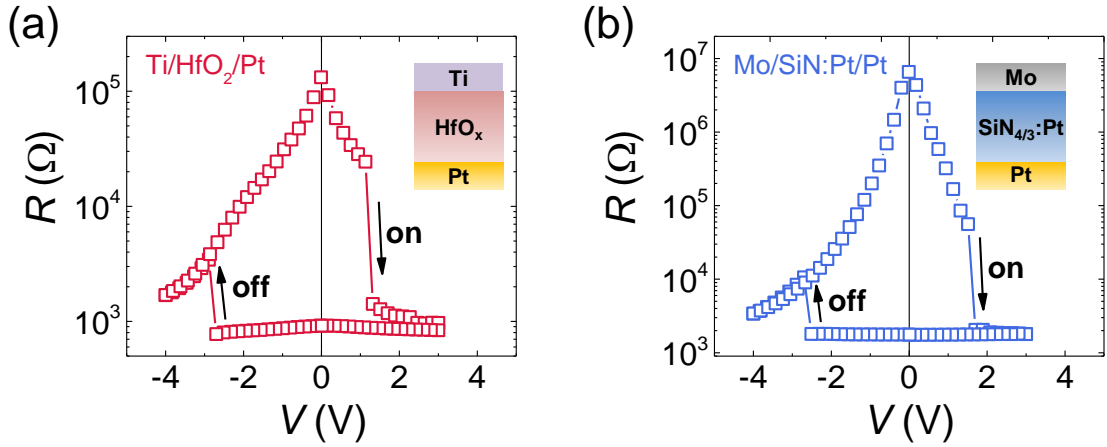

**Figure 1.** Bipolar resistance switching curves in **(a)** filamentary Ti/HfO$_2$/Pt device (forming not shown) and **(b)** nanometallic Mo/SiN$_{3/4}$:9.3%Pt/Pt device, both with cell radius of 80 μm. Inset: schematics of materials stacks. Arrows indicate switching directions.

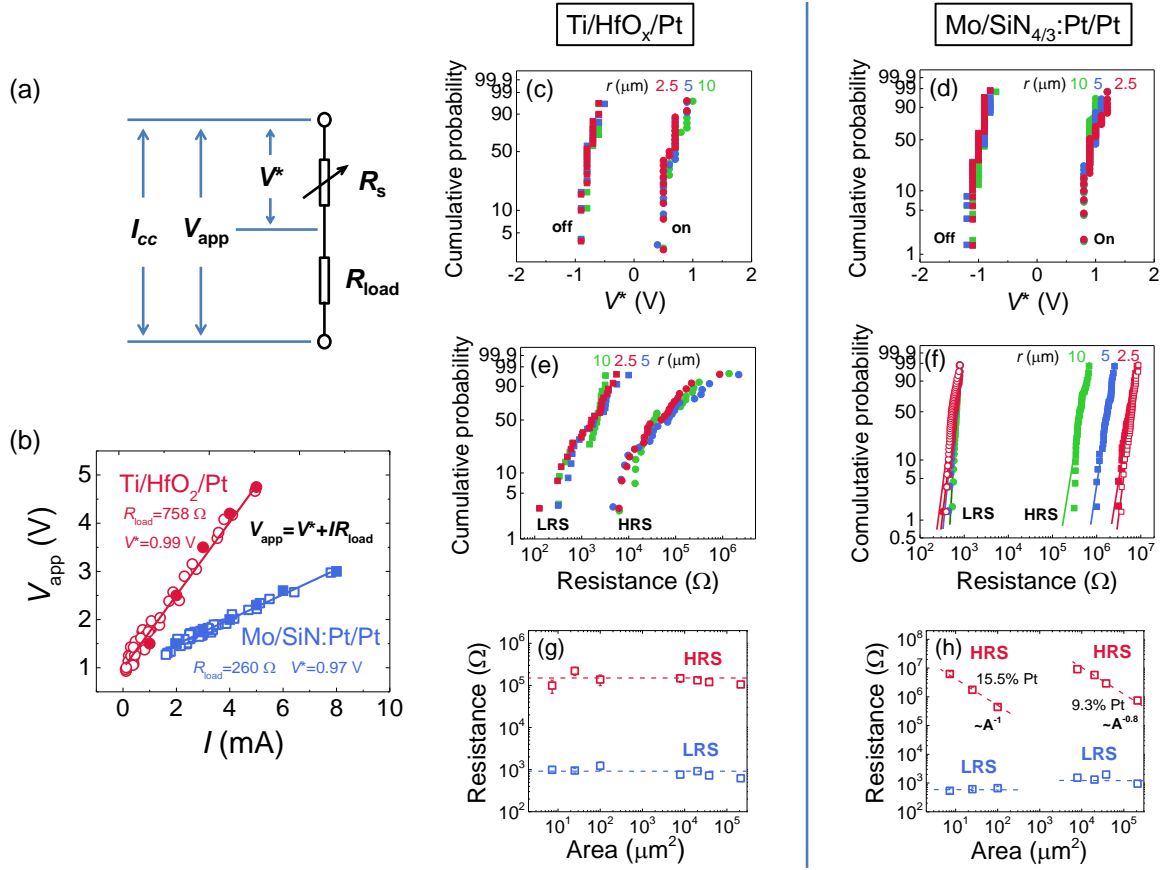

**Figure 2.** Voltage-controlled resistance switching in RRAMs of different areas. **(a)** Equivalent circuit for RRAMs: switching element $R_S$ with voltage $V^*$ in serial connection with load resistor $R_{load}$, which mostly comes from electrode's spreading resistance. **(b)** Applied voltage $V_{app}$ vs. $I$, at switching, which defines $R_{LRS/HRS} = V_{app}/I$ in off-switching (empty symbols) and on-switching (filled symbols) for Ti/HfO$_2$/Pt (red) and Mo/SiN$_{3/4}$:15.5%Pt/Pt (blue). Data include devices of radius from 10 μm to 2.5 μm. Solid lines: linear correlation consistent with $V_{app}=V^* + IR_{load}$. **(c)-(f)** Weibull statistics of switching voltage and resistance in Ti/HfO$_2$/Pt RRAM (left panel) and Mo/SiN$_{3/4}$:15.5%Pt/Pt (right panel). While $V^*$ is uniform in both RRAMs, resistance for HRS and LRS is size-independent and more scattered in Ti/HfO$_2$/Pt. In contrast, in Mo/SiN$_{3/4}$:Pt/Pt LRS but not HRS is area independent, which include both DC switching (filled symbols) and 100 ns pulse switching (empty symbols) data. **(g)-(h)** Summary of area dependence of resistance.

**Table 1.** Summary of literature $V_{app}$-$I$ data, replotted in linear-linear scale, of various RRAM devices (ref. 30-40, reference number indicated in brackets in plots). Critical voltage $V^*$ (the intercept) and circuit load $R_{load}$ (the slope) extracted from linear plot according to Eq. (1) are listed on the right.

| $V_{app}$-$I$ | $V^*$ (V) | $R_{load}$ (Ω) |
|---|---|---|
| 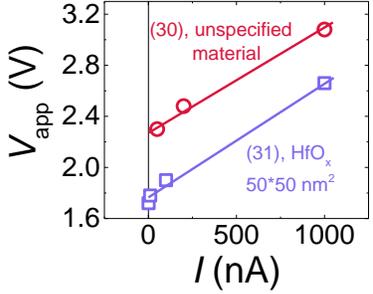 | 2.27 | 800k |
| | 1.76 | 900k |
| 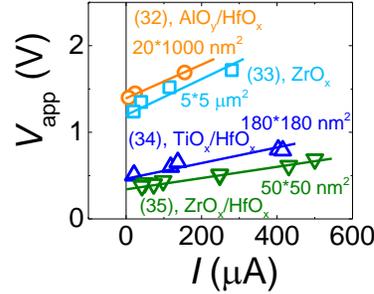 | 1.4 | 1900 |
| | 1.26 | 1700 |
| | 0.51 | 717 |
| | 0.36 | 609 |
| 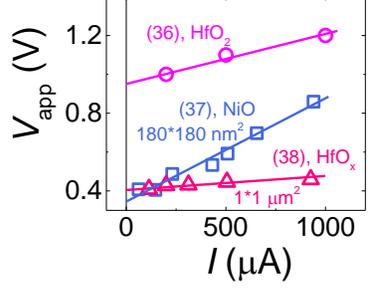 | 0.96 | 245 |
| | 0.34 | 525 |
| | 0.41 | 55 |
| 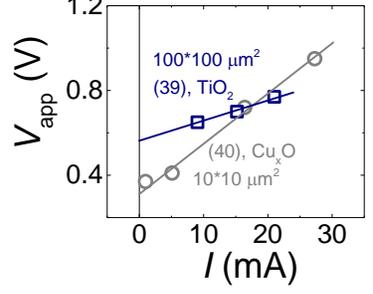 | 0.55 | 10 |
| | 0.33 | 23 |

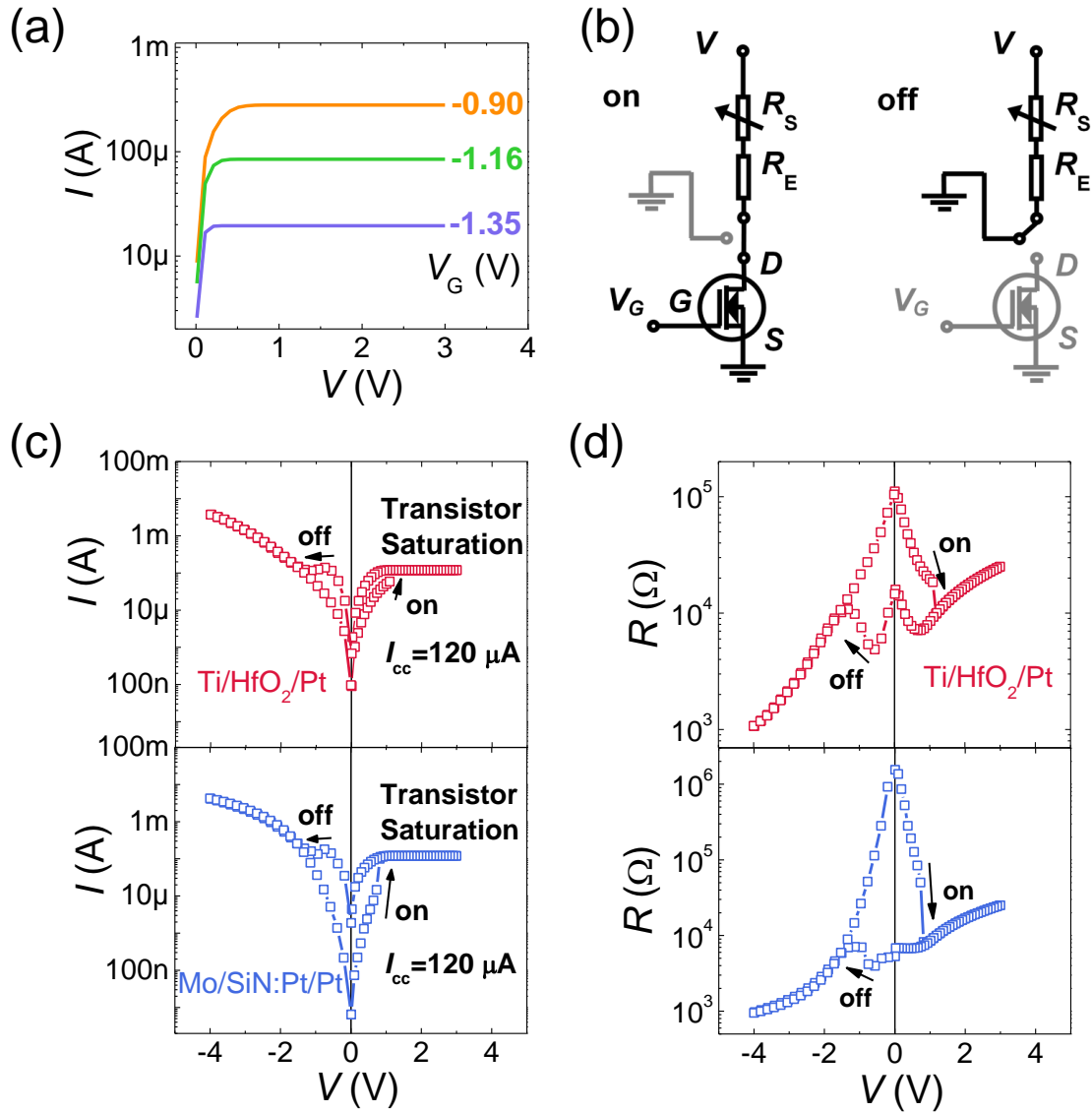

**Figure 3.** (a) Characteristic *I-V* curves of transistor with gate-voltage ($V_G$)-tunable saturation current. (b) Schematic of measurement set-up using transistor (T) to supply current compliance on RRAM (comprised of switching element $R_S$ and electrode load $R_E$) during on-switching, and no compliance during off-switching. (c) *I-V* switching curves of Ti/HfO$_2$/Pt and Mo/SiN$_{3/4}$:9.3%Pt/Pt devices with on-current limited by transistor's saturation current. Arrows indicate on- and off-switching. (d) Same data as (c) plotted in terms of resistance $R$ (=$I/V$) vs. *V*.

**Table 2.** Summary of materials, switching properties and scaling behaviors of Ti/HfO$_2$/Pt and Mo/SiN$_{4/3}$:Pt/Pt RRAM devices. $R$, $C$, $I$ and $V^*$ represent resistance, capacitance, switching current and critical switching voltage on $R_S$, respectively. Experimental data are shown in Figure 2, Figure 4, Figure S5, Ref. 11, Ref. 23, Ref. 24 and Ref. 26.

| *Type* / *Characteristics* | | **Filamentary RRAM** | | **Nanometallic RRAM** | |
|---|---|---|---|---|---|
| MIM Stack | | Ti/HfO$_2$/Pt | | Mo/SiN$_{4/3}$:Pt/Pt | |
| Film Condition | | 10 nm HfO$_x$ | | 7.5 nm SiN$_{4/3}$:9.3%Pt ($r$ > 40 μm) <br> 2.5 nm SiN$_{4/3}$:15.5%Pt ($r$ < 10 μm) | |
| Deposition Method | | Thermal ALD (250°C) | | Co-sputtering | |
| $R$ | States | HRS | LRS | HRS | LRS |
| | Apparent Area ($A$) dependence | $R \sim$ const. (Fig. 2g&S5) | Constant $I_{cc}$: $R \sim$ const. (Fig. S5c) <br> Constant $J_{cc}$: $R \sim 1/A$ (Fig. S5a) | $R \sim 1/A$ (Fig. 2h&S6) | Constant $I_{cc}$: $R \sim$ const. (Fig. S5d) <br> Constant $J_{cc}$: $R \sim 1/A$ (Fig. S5b) |
| | Temperature ($T$) dependence | d$R$/d$T$ < 0 | d$R$/d$T$ > 0 (Ref. 26) | d$R$/d$T$ < 0 | d$R$/d$T$ > 0 (Ref. 24) |
| $C$ | $R$, $A$ dependence | $C \sim$ const., $C \sim A$ (Ref. 11) | | $C \sim R$, $C \sim A$ (Ref. 23) | |
| $I$ | | $I_{on} = I_{off}$ (Fig. 4a) | | $I_{on} = I_{off}$ (Fig. 4b) | |
| $V^*$ | on | + 1.31 ± 0.15 V (Fig. 4c) | | + 1.13 ± 0.10 V (Fig. 4c) | |
| | off | − 0.85 ± 0.09 V (Fig. 4d) | | − 0.92 ± 0.08 V (Fig. 4d) | |

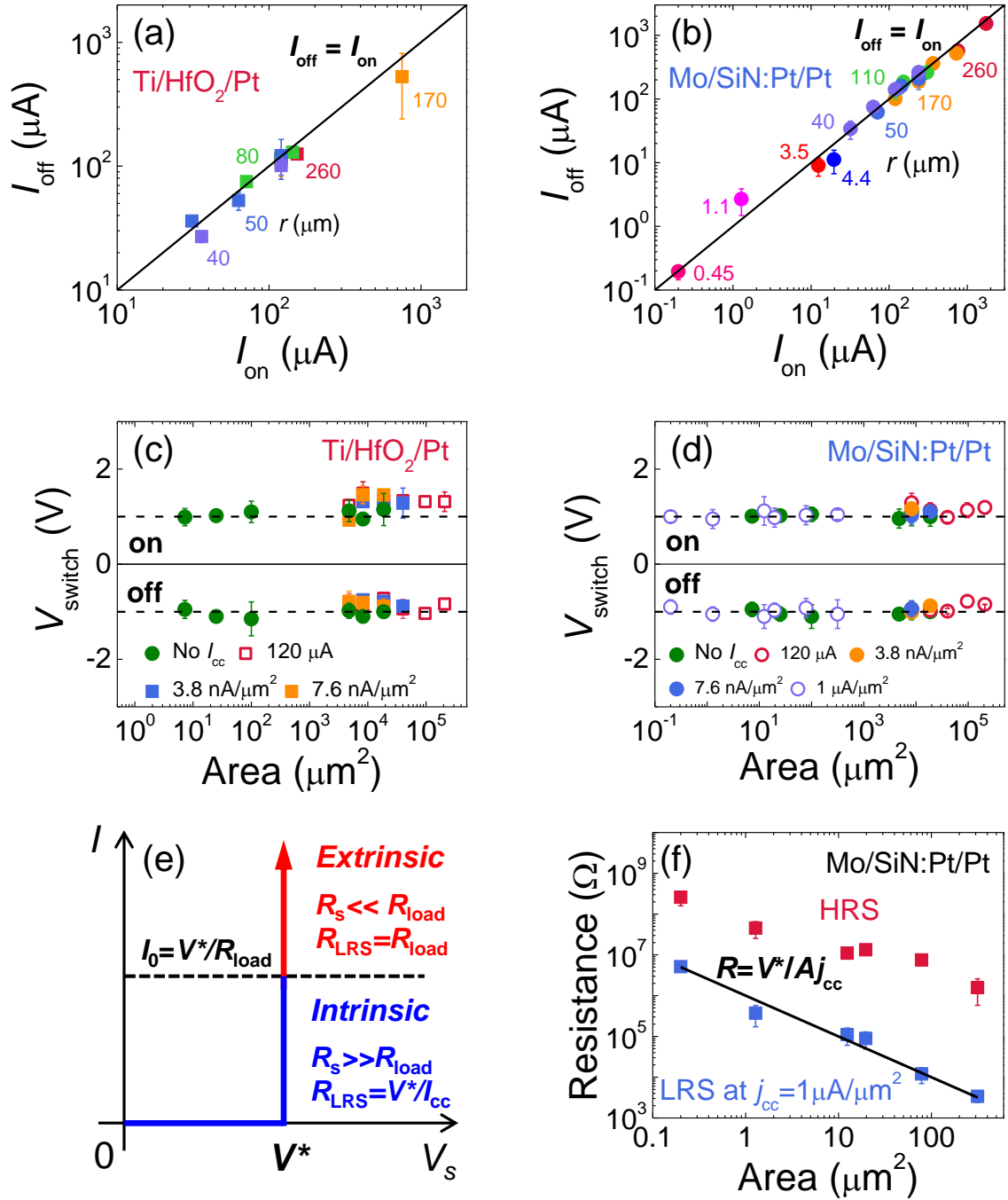

**Figure 4.** Voltage-controlled resistance switching and area scalability. Correlation between on-switching current ($I_{on}=I_{cc}$) and off-switching current ($I_{off}$) for **(a)** Ti/HfO$_2$/Pt and **(b)** Mo/SiN$_{3/4}$:Pt/Pt RRAM, of various sizes (radius labeled). Identical $I_{off}$ and $I_{cc}$ set in on-switching indicate same on/off voltage regardless of area and current level. **(c-d)** On/off voltages on $R_S$ are constant regardless of area, compliance current and compliance current density for Ti/HfO$_2$/Pt and Mo/SiN$_{3/4}$:9.3%Pt/Pt RRAM. Error bars in (a)-(d) are from the average of 10 tests. **(e)** Schematic *I-V* curve in switching element $R_S$ (its voltage $V_S$) during on-switching, with total internal and external load equal to $R_{load}$ and $V^*$ being characteristic on-switching voltage. In intrinsic regime, $R_S \ll R_{load}$, so LRS resistance ($R_{LRS}$) is essentially $R_{load}$ without any area dependence. In extrinsic regime, $R_S \gg R_{load}$, so $R_{LRS}$ is approximately $V^*/I_C$, thus becoming area-dependent when a constant current density $j_{cc}$ is applied ($R_{LRS}=V^*/Aj_{cc}$, $A$ being the apparent device area). **(f)** Area scaling of HRS and LRS resistance for Mo/SiN$_{4/3}$:15.5%Pt/Pt RRAM devices. Solid line is predicted LRS resistance, $V^*/Aj_{cc}$, with $V^*$=1 V and $j_{cc}$=1 µA/µm$^2$. LRS resistance of 1 MΩ was obtained in 0.2 µm$^2$ device with off/on ratio larger than 100. Error bars are from average of 10 tests.

## ASSOCIATED CONTENT

**Supporting Information**.

The Supporting Information is available free of charge.

Supporting results: Figs S1-S8.

## AUTHOR INFORMATION

**Corresponding Author**

*Email: iweichen@seas.upenn.edu.
*Email: iweichen@seas.upenn.edu.

**Author Contributions**

The manuscript was written through contributions of all authors. All authors have given approval to the final version of the manuscript. ‡These authors contributed equally.

## ACKNOWLEDGMENTS

This research was supported by the US National Science Foundation Grant No. DMR-1409114. The use of facilities at Penn's LRSM supported by DMR-1120901 is gratefully acknowledged.
This research was supported by the US National Science Foundation Grant No. DMR-1409114. The use of facilities at Penn's LRSM supported by DMR-1120901 is gratefully acknowledged.

REFERENCES

1. R. Waser, R. Dittmann, G. Staikov, K. Szot, *Adv. Mater.* **2009**, *21*, 2632-2663.

2. J. J. Yang, D. B. Strukov, D. R. Stewart, *Nature Nanotech.* **2013**, *8*, 13-24.

3. S. H. Jo, T. Chang, I. Ebong, B. B. Bhadviya, P. Mazumder, W. Lu, *Nano Lett.* **2010**, *10*, 1297-1301.

4. C. S. Hwang, *Adv. Electron. Mater.* **2015**, *1*(6).

5. D. S. Jeong, K. M. Kim, S. Kim, B. J. Choi, C. S. Hwang, *Adv. Electron. Mater.* **2016**, 2 (9).

6. H. S. P. Wong, H. Y. Lee, S. Yu, Y. S. Chen, Y. Wu, P. S. Chen, B. Lee, F. T. Chen, M. J. Tsai, *Proceedings of the IEEE* **2012**, *100*, 1951-1970.

7. Valov, R. Waser, J. R. Jameson, M. N. Kozicki, *Nanotechnology* **2011**, *22*, 254003.

8. H. Y. Peng, G. P. Li, J. Y. Ye, Z. P. Wei, Z. Zhang, D. D. Wang, G. Z. Xing, T. Wu, *Appl. Phys. Lett.* **2010**, *96*, 192113.

9. H. D. Kim, H. M. An, E. B. Lee, T. G. Kim, *IEEE Trans. Electron Devices* **2011**, *58*, 3566-3573.

10. B. Govoreanu, G. S. Kar, Y.-Y. Chen, V. Paraschiv, S. Kubicek, A. Fantini, I. P. Radu, L. Goux, S. Clima, R. Degraeve, N. Jossart, O. Richard, T. Vandeweyer, K. Seo, P. Hendrickx, G. Pourtois, H. Bender, L. Altimime, D. J. Wouters, J. A. Kittl and M. Jurczak, in *IEDM Tech. Dig*, **2011**, pp. 729-732.

11. Y. Lu, J. H. Lee, I. W. Chen, *ACS nano* **2015**, *9*, 7649-7660.

12. S.-E. Ahn, M.-J. Lee, Y. Park, B. S. Kang, C. B. Lee, K. H. Kim, S. Seo, D.-S. Suh, D.-C. Kim, J. Hur, W. Xianyu, G. Stefanovich, H. Yin, I.-K. Yoo, J.-H. Lee, J.-B. Park, I.-G. Baek, B. H. Park, *Adv. Mater.* **2008**, *20*, 924-928.

13. B. Govoreanu, A. Redolfi, L. Zhang, C. Adelmann, M. Popovici, S. Clima, H. Hody, V. Paraschiv, I.P. Radu, A. Franquet, J.-C. Liu, J. Swerts, O. Richard, H. Bender, L. Altimime M. Jurczak, in *IEDM Tech. Dig*, **2013**, pp. 10-2.

14. D. Ielmini, F. Nardi, C. Cagli, *IEEE Electron Device Lett.* **2011**, *58*, 3246-3253.

15. D. Ielmini, *IEEE Trans. Electron Devices* **2011**, *58*, 4309-4317.

16. M. Kund, G. Beitel, C. U. Pinnow, T. Rohr, J. Schumann, R. Symanczyk, K. D. Ufert, G. Muller, *IEDM Tech. Dig*, **2005**, pp. 5-5.

17. F. Nardi, D. Ielmini, C. Cagli, S. Spiga, M. Fanciulli, L. Goux, D. J. Wouters, *Solid-State Electron.* **2011,** *58*, 42-47.

18. S. Tappertzhofen, I. Valov, R. Waser, *Nanotechnology* **2012**, *23*, 145703.

19. Y. Yang, W. Lu, *Nanoscale* **2013**, *5*, 10076-10092.

20. B. Chen, S. G. Kim, Y. Wang, W.-S. Tung, I. W. Chen, *Nature Nanotech.* **2011**, *6*, 237-241.

21. J. Choi, A. B. Chen, X. Yang, I. W. Chen, *Adv. Mater.* **2011**, *23*, 3847-3852.

22. Y. Lu, J. H. Lee, X. Yang, I. W. Chen, *Nanoscale* **2016**, *8*, 18113-18120.

23. X. Yang, I. Tudosa, B. J. Choi, A. B. Chen, I. W. Chen, *Nano Lett.* **2014**, *14*, 5058-5067.

24. X. Yang, *Resistance switching devices based on amorphous insulator-metal thin films*. Ph.D. Dissertation, University of Pennsylvania, **2014**, arXiv:1412.2083.

25. J. Y. Seok, S. J. Song, J. H. Yoon, K. J. Yoon, T. H. Park, D. E. Kwon, H. Lim, G. H. Kim, D. S. Jeng, C. S. Hwang, *Adv. Funct. Mater.* **2014**, *24*, 5316-5339.

26. Y. Lu, I. W. Chen, **2017**, arXiv preprint arXiv:1703.02003.

27. A. B. Chen, B. J. Choi, X. Yang, I. W. Chen, *Adv. Funct. Mater.* **2012**, *22*, 546-554 .

28. X. Yang, I. W. Chen, *Sci. Rep.* **2012**, *2*, 744.

29. X. Yang, A. B. Chen, B. J. Choi, I. W. Chen, *Appl. Phys. Lett.* **2013**, *102*, 043502.

30. S. G. Park, et al. *IEDM Tech. Dig*, **2012**, pp. 20-8.

31. Q. Luo, et al. *IEDM Tech. Dig*, **2015**, pp. 10-2.

32. B. Gao, et al. **2014**, *ACS nano*, *8*, 6998-7004.

33. M. C. Wu, Y. W. Lin, W. Y. Jang, C. H. Lin, T. Y. Tseng, *IEEE Electron Device Lett.* **2011**, *32*, 1026-1028.

34. H. Y. Lee, et al. *IEDM Tech. Dig*, **2008**, pp. 1-4.

35. J. Lee, *IEDM Tech. Dig*, **2010**, pp. 19-5.

36. E. Vianello, *IEDM Tech. Dig*, **2014**, pp. 6-3.

37. K. Tsunoda, et al. *IEDM Tech. Dig*, **2007**, pp. 767-770.

38. D. Ielmini, F. Nardi, S. Balatti, *IEEE Trans. Electron Devices*, **2012**, *59*, 2049-2056.

39. X. Cao, X. M. Li, X. D. Gao, Y. W. Zhang, X. J. Liu, Q. Wang, L. D. Chen, *Appl. Phys.* **2009**, *97*, 883-887.

40. S. Y. Wang, C. W. Huang, D. Y. Lee, T. Y. Tseng, T. C. Chang, *J. Appl. Phys.* **2010**, *108*, 114110.

41. S. Yu, X. Guan, H. S. P. Wong, *IEDM Tech. Dig*, **2011**, pp. 17-3.

42. N. Raghavan, R. Degraeve, A. Fantini, L. Goux, D. J. Wouters, G. Groeseneken, M. Jurczak, *IEDM Tech. Dig*, **2013**, pp. 21-1.

43. M. J. Lee, *et al*. *Nature Mater.* **2011**, *10*, 625-630.

44. P. Huang, *et al*. *IEEE Trans. Electron Devices* **2013**, *60*, 4090-4097.

# Supporting Information

**Scalability of Voltage-Controlled Filamentary and Nanometallic Resistance Memories**

*Yang Lu[‡], Jong Ho Lee[‡], and I-Wei Chen\**

[‡]These two authors contribute equally to this work.

Department of Materials Science and Engineering, University of Pennsylvania, Philadelphia, PA 19104-6272, USA

E-mail: iweichen@seas.upenn.edu

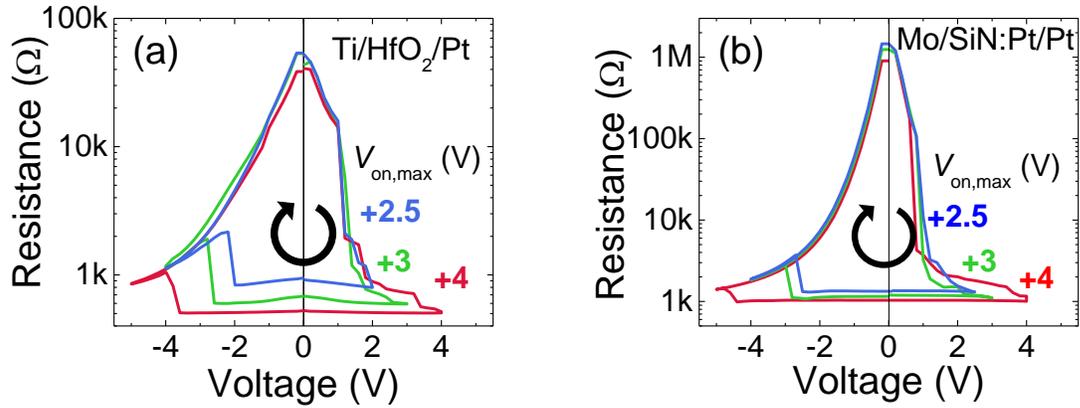

**Figure S1. Same-cell *R*-*V* switching curves with different maximum ± sweep voltage ($V_{on,max}$ indicated for positive polarity) for (a) Ti/HfO$_2$/Pt RRAM, and (b) Mo/SiN$_{4/3}$:Pt/Pt RRAM.** Cell radius: 80 μm. In both cases, a larger current during the on-switching positive sweep leads to a progressive decrease of LRS and increase of off-switching voltage. $R_{LRS}$ approaches an asymptote ($R_S+R_{load}$). On-switching always occurs around a critical voltage $V^*$ of +1 V, as further verified in Fig. 2b.

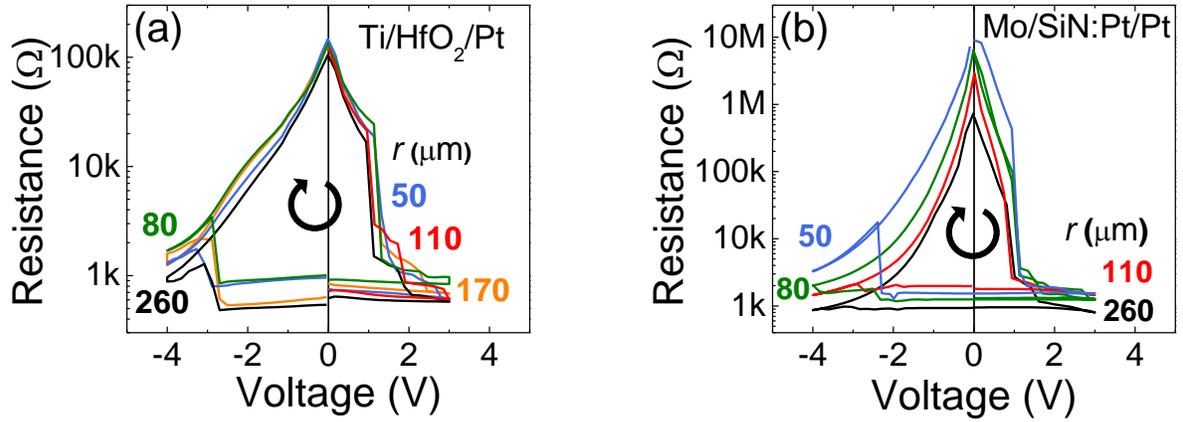

**Figure S2. Area scaling of resistance switching in Ti/HfO$_2$/Pt RRAM and Mo/SiN$_{4/3}$:Pt/Pt RRAM.** *R-V* switching curves of different device radius *r* of **(a)** Pt/HfO$_2$/Ti RRAM and **(b)** Mo/SiN$_{4/3}$:9.3%Pt/Pt RRAM. On-switching occurs at a size-independent critical voltage *V*\* of +1 V in both RRAMs. Size-independent HRS and LRS resistance of Ti/HfO$_2$/Pt, and size-independent LRS but size-dependent HRS of Mo/SiN$_{4/3}$:Pt/Pt, are summarized in Fig. 2g-h.

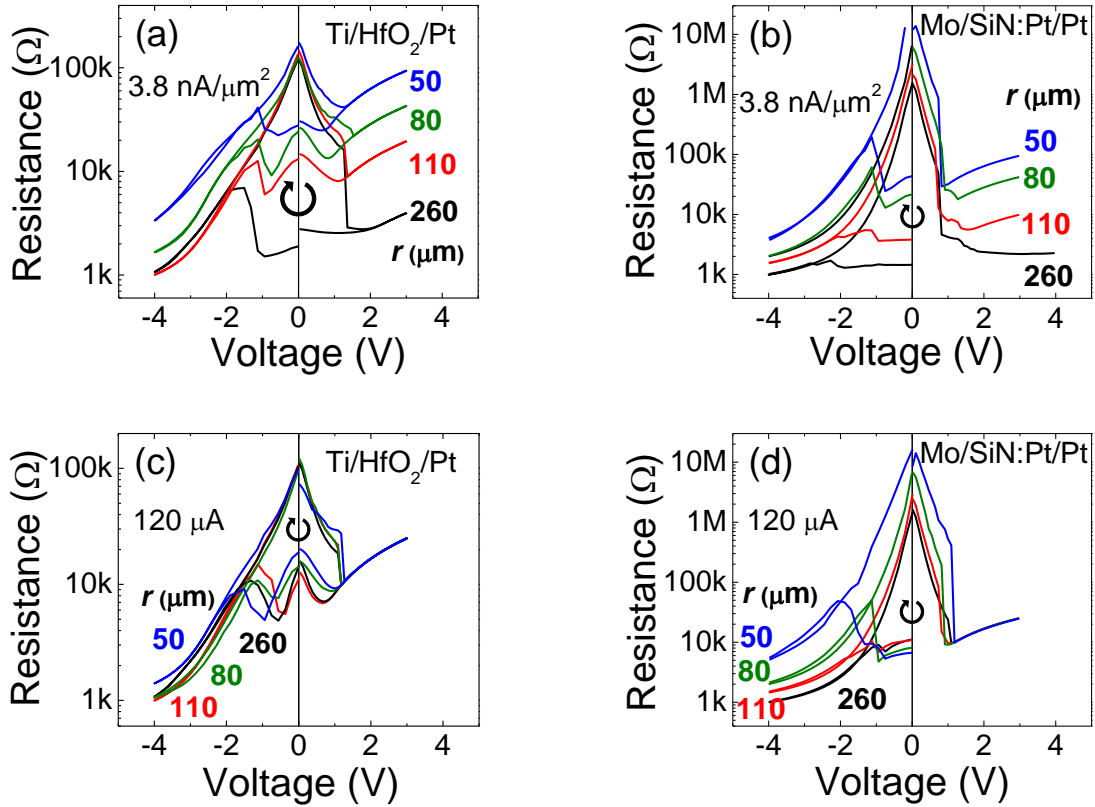

**Figure S3. *R-V* curves of different device radius *r* under compliance control. (a)** Ti/HfO$_2$/Pt RRAM and **(b)** Mo/Si$_3$N$_4$:9.3%Pt/Pt RRAM, with same constant current density 3.8 nA/μm$^2$ set for on-switching, resulting in area-dependent LRS in both. **(c)** Ti/HfO$_2$/Pt RRAM **(d)** Mo/Si$_3$N$_4$:9.3%Pt/Pt RRAM, with same constant current 120 μA set for on-switching, resulting in area-independent LRS in both. Under both sets of compliance, zero-voltage HRS is area-dependent in Mo/Si$_3$N$_4$:9.3%Pt/Pt (b, d), but area-independent in Ti/HfO$_2$/Pt (a, c).

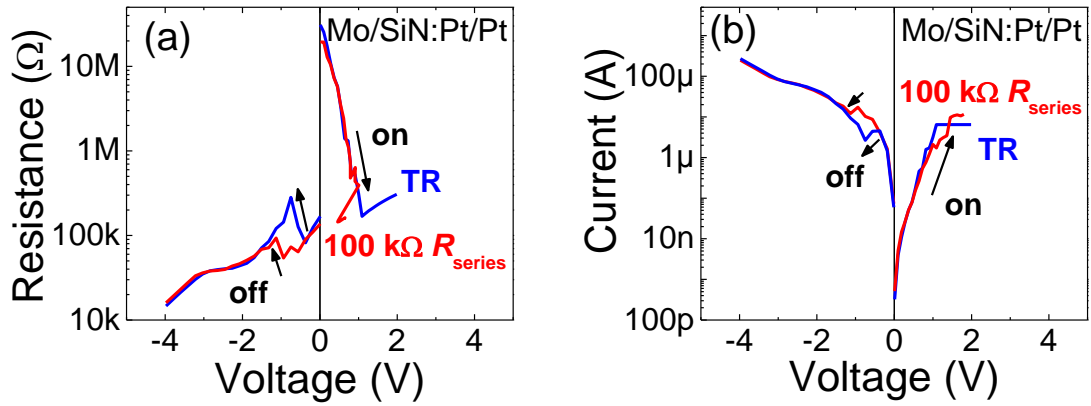

**Figure S4. Comparison of switching curves of Mo/Si$_3$N$_4$:9.3%Pt/Pt RRAM in two configurations of compliance control, with a transistor (TR), and with a 100 kΩ resistor ($R_{series}$). (a)** Resistance *vs*. voltage on the RRAM device. **(b)** Current *vs*. voltage applied to the RRAM device. Arrows indicate on- and off-switching. Voltage sharing on $R_{series}$ causes a higher off-switching voltage in both (a) and (b).

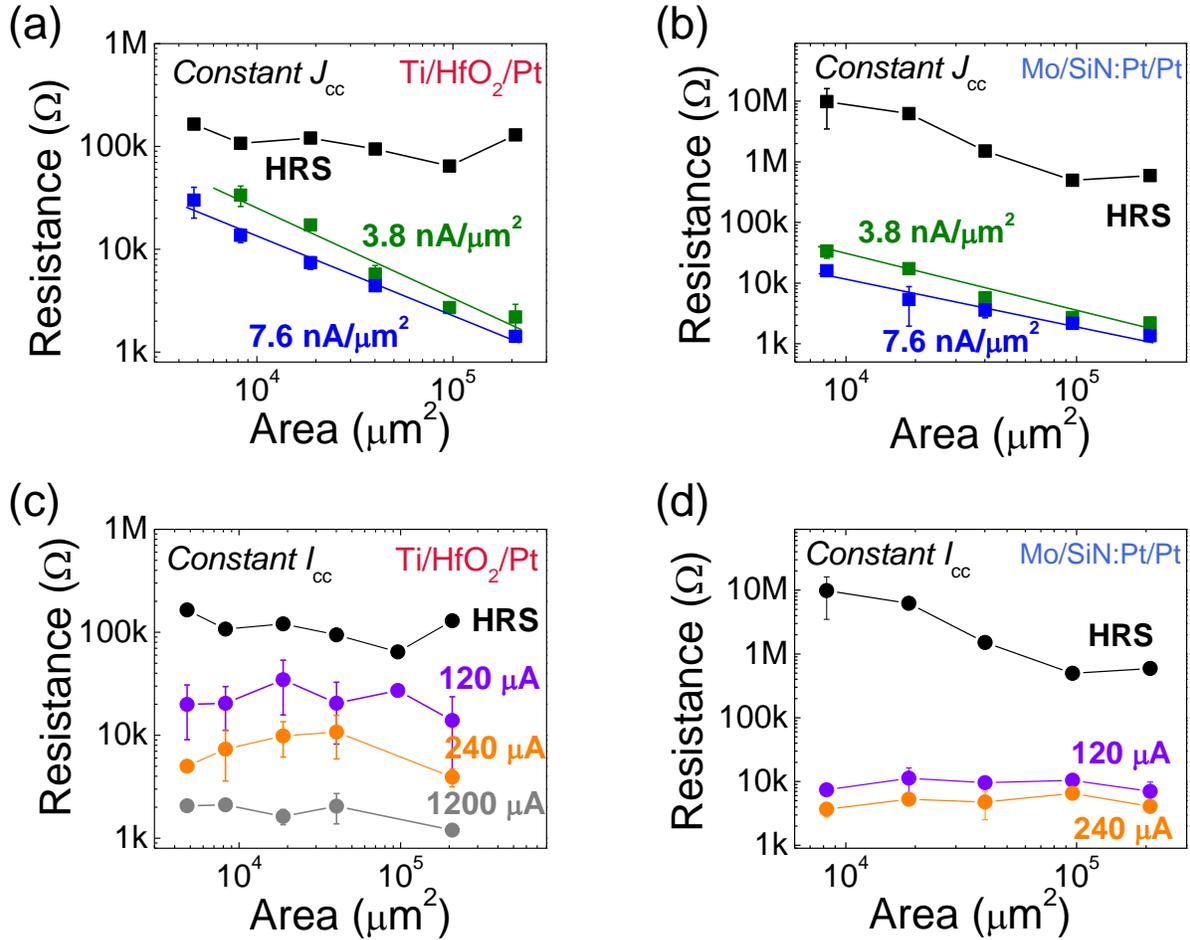

**Figure S5. Resistance scaling of Ti/HfO$_2$/Pt RRAM and Mo/SiN$_{4/3}$:Pt/Pt RRAM under current compliance.** **(a)** and **(c)**: Ti/HfO$_2$/Pt device with area-independent HRS has area-dependent LRS under the same current density (a), but area-independent LRS under the same current (c). **(b)** and **(d)**: Mo/SiN$_{4/3}$:9.3%Pt/Pt device with area-dependent HRS has area-dependent LRS under the same current density (b), but area-independent LRS under the same current (d). Error bars are from average of 10 tests. Current and current density control in 1T1R configuration. Straight lines next to LRS data in (a) and (b): predicted $R_{LRS}=V^*/Aj_{cc}$ similar to the one in Fig. 4f.

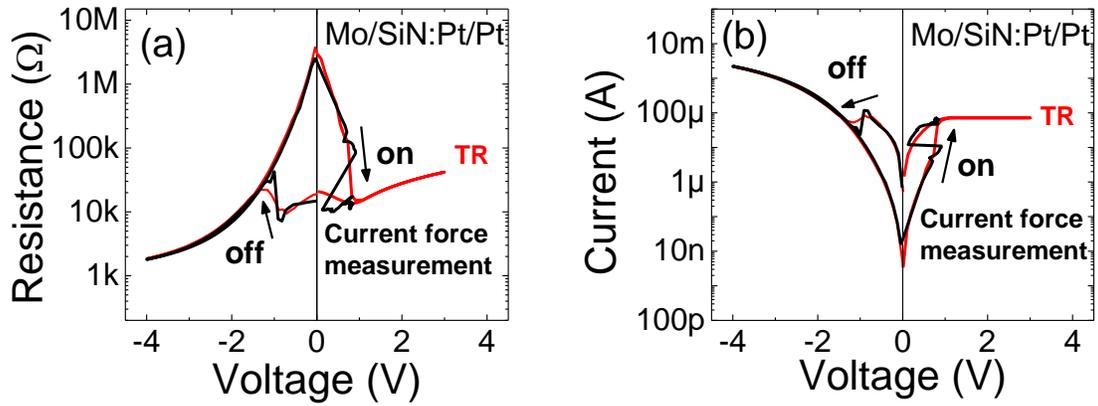

**Figure S6. Comparison of switching curves of Mo/SiN$_{4/3}$:9.3%Pt/Pt RRAM in voltage sweep with compliance provided by transistor (TR, red), and in current (force) sweep, with current ramping from 0 to 70 μA under a positive bias causing on-switching (black curves under negative bias).** The red curves under negative bias for off-switching were obtained in voltage sweep without transistor. **(a)** Resistance *vs*. voltage curves and **(b)** current *vs*. voltage curves. Arrows indicate switching directions. In current sweep, a constant onset voltage around 1 V triggers on-switching that results in a voltage drop as well as a resistance drop.

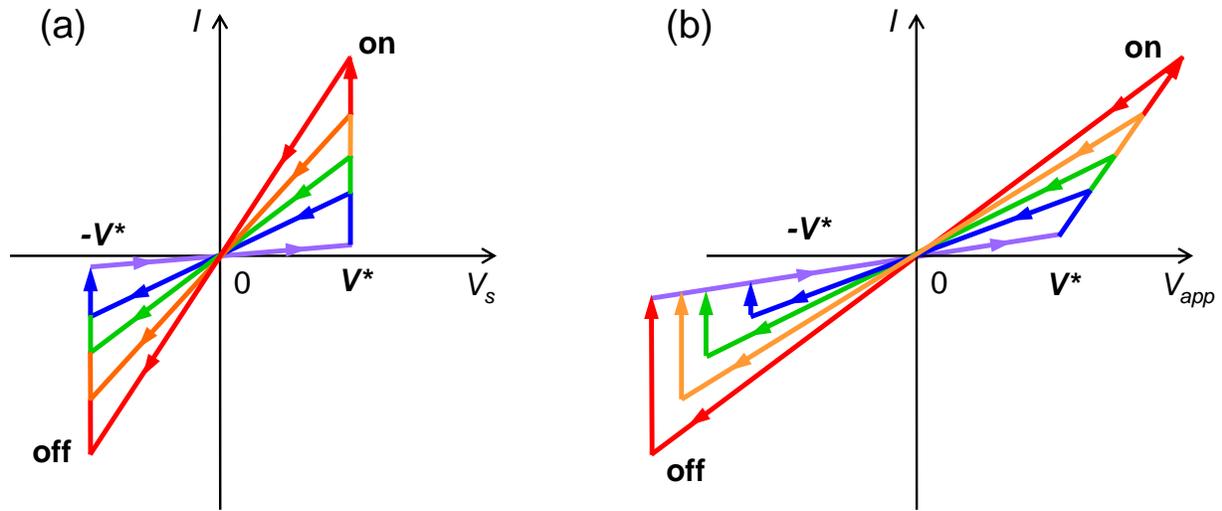

**Figure S7. Schematic *I-V* curves of voltage-controlled bipolar resistance switching.** **(a)** Current (*I*) vs voltage on $R_s$ ($V_s$): On- and off-switching occur at $V_s = \pm V^*$ to reach certain current *I*. **(b)** Current (*I*) vs total applied voltage $V_{app}$. Unlike (a), on- and off-switching occur at various voltages because of load sharing, but maximum $V_{app}$ is the same in both polarity. As $V_{app}$ increases during on-switching, *I-V* curve of switched LRS asymptotically approaches $I=V_{app}/R_{load}$. Arrows indicate sweeping directions. Refer to equivalent circuit in Fig. 2a.

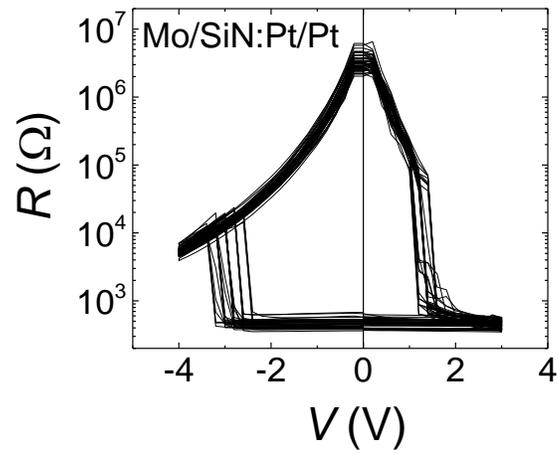

**Figure S8.** Overlapping of 100 cycling *R-V* switching curves of Mo/SiN$_{4/3}$:Pt/Pt device with radius of 2.5 μm show good cycle-to-cycle uniformity of switching characteristics.